\documentclass[12pt,a4paper]{article}

\usepackage{amsmath,amssymb}

\newcommand{\bc}{\begin{center}}
\newcommand{\ec}{\end{center}}
\newcommand{\be}{\begin{equation}}
\newcommand{\ee}{\end{equation}}
\newcommand{\ba}{\begin{array}}
\newcommand{\ea}{\end{array}}
\newcommand{\bea}{\begin{eqnarray}}
\newcommand{\eea}{\end{eqnarray}}
\newcommand{\bt}{\begin{tabular}}
\newcommand{\et}{\end{tabular}}

\newcommand{\bsl}{\boldsymbol}

\begin{document}
\title{\bf $P$ Wave Baryons in Field Correlator Method: Hyperons}

\author{O.~N.~Driga,
I.~M.~Narodetskii and A.~I.~Veselov \\
{\itshape Institute of Theoretical and Experimental Physics,
117218 Moscow}}

\maketitle \vspace{1cm}

\begin{abstract}\noindent
We provide an investigation of the $P$-wave hyperons employing the
field correlator method in QCD. This method allows to derive the
Effective Hamiltonian successfully applied to the meson and ground
state baryon spectra. The hyperon spectrum appears to be expressed
through two parameters relevant to QCD, the string tension
$\sigma$, the strong coupling constant $\alpha_s$, and the bare
strange quark mass $m_s$. Using these parameters a unified
description of the ground and excited hyperon states is achieved.
We also briefly consider the nucleon $P$-wave excitations. In
particular, we predict that both the nucleon and hyperon states
have the similar cost (in $\Delta L$) $\sim$ 460 MeV.

\end{abstract}

\section{Introduction}

The advent of new ideas concerning quark-quark forces in QCD have
led to revival of interest in baryon spectroscopy. The
spectroscopy of heavy baryons has undergone a great renaissance in
recent years, providing an exceptional window into tests of QCD,
see {\it e.g.} \cite{R06}. As to the ``old'' $\Xi$ resonances,
which we consider in this paper, nothing has changed since 1996.
In the last Particle Data Group review \cite{PDG06} five
resonances remain stuck with the same one star $^*$ or two stars
$^{**}$, meaning doubtful \footnote{The ambitious program of $\Xi$
spectroscopy has been proposed at JLab \cite{Guo}, but we do not
have any finished results so far.}. Most of the spin-parity values
of $\Xi$ have not been measured but have been assigned in accord
with expectation of the theory. Besides, knowledge of excited
states is very much limited. Therefore, a powerful guideline for
assigning quantum numbers to new states is required both by theory
and experiment.

The reproduction of the baryon mass spectrum from first principles
is an important challenge for QCD. Ground state spectroscopy on
the lattice is by now a well understood problem and impressive
agreement with experiments has been achieved. However, the lattice
study of excited states is not so advanced, see \cite{B06} and
references therein. The purpose of this paper is to present a
consistent treatment  of the $P$-wave hyperons within the
alternative method in QCD, the field correlator method (FCM)
\cite{DS}. The similar analysis of heavy $c\,-$ and $b\,-$ baryons
will be given elsewhere \cite{NV}.

In the FCM one derives the Effective Hamiltonian (EH), which
comprises both confinement and relativistic effects, and contains
only universal parameters: the string tension $\sigma$, the strong
coupling constant $\alpha_s$, and the bare (current) quark masses.
The simple local form of this Hamiltonian occurs for the objects
with temporal scale larger than the vacuum gluon correlation scale
$T_g\,\sim\,0.2$ fm, {\it i.e.} it is applicable to all states,
perhaps with an exception of bottomonium. There is a lot of
calculations of masses and wave functions of light mesons
heavy quarkonia
and heavy-light mesons \cite{KNS}, but only a few for S-wave
baryons  \cite{NT}. The present investigation was initially
motivated as an attempt to extend an approach of Refs. \cite{NT}
for the $P\,-\,$ wave low-lying orbitally excited baryons. As in
Ref. \cite{NT}, we compute only the confinement energies
(corrected by the perurbative one gluon exchange potential) and
disregard the spin-spin and spin-orbit interactions.

The paper is organized as follows. In Section 2 we briefly review
the EH method. In Section 3 we consider the hyperspherical
approach which is a very effective numerical tool to solve the EH.
In Section 4 our predictions for the P-wave nucleons and strange
baryons are reported and compared with the results of other
approaches. Section 5 contains our conclusions.

\section{The Effective Hamiltonian in  FCM}The
application of the method for the baryons was described in detail
elsewhere \cite{NT}. Here we give only a brief summary important
for this particular calculation.

The EH has the following form:
\begin{equation}
\label{eq:H} H=\sum\limits_{i=1}^3\left(\frac{{ m}_{i}^2}{2\mu_i}+
\frac{\mu_i}{2}\right)+H_0+V.
\end{equation}
Here $H_0$ is the nonrelativistic kinetic energy operator and $V$
is the sum of the string potential $V_{\rm Y}({\bf r}_1,\,{\bf
r}_2,\,{\bf r}_3)$ and a Coulomb interaction term arising from one
gluon exchange. The string potential considered in this work is
\be V_{Y}({\bf r}_1,\,{\bf r}_2,\,{\bf
r}_3)\,=\,\sigma\,r_{min},\ee where $\sigma$ is the string tension
and  $r_{min}$ is the minimal length string corresponding to the
Y-~shaped configuration. In this picture, strings start from each
quark and  meet at the Toricelli point of the triangle formed by
the three quarks. This point is such that it minimizes the sum of
the string lengths, and its position is a complicated function of
the quark coordinates $\bf{r}_{i}$.
The Coulomb interaction is \be V_{\rm
Coulomb}=-\frac{2}{3}\,\alpha_s\,\sum\limits_{i<j}\frac{1}{r_{ij}},
\ee where $\alpha_s$ is the strong coupling constant and $r_{ij}$
are the distances between quarks.

In Eq. (\ref{eq:H}) ${m}_{i}$ are the bare quark masses, while
$\mu_i$ are the constant auxiliary einbein fields, initially
introduced  in order to get rid of the square roots appearing in
the relativistic Hamiltonian \cite{DKS}.
The dynamics remains essentially relativistic, though being
non-relativistic in form.
The einbein fields are eventually treated as variational
parameters. The eigenvalue problem is solved for each set of
$\mu_i$, then one has to minimize $\langle H\rangle$ with respect
to $\mu_i$. Such an approach allows one a very transparent
interpretation of einbeins:  $\mu_i$ can be treated as constituent
masses of quarks of current mass ${m}_i$. In this way the notion
of constituent masses
arises.

The baryon mass is then given by formula
\begin{equation}
\label{M_B}
 M_B\,=\,\sum\limits_{i=1}^3\left(\frac{{ m}_{i}^2}{2\mu_i}+
\frac{\mu_i}{2}\right)\,+\,E_0(m_i,\mu_i)
\, +\,C,
\end{equation}
where $E_0(m_i,\mu_i)$ is an eigenvalue of the operator
$H_0\,+\,V,$ ~$\mu_i$ are defined from the condition
\begin{equation}
\label{partial}
\frac{\partial}{\partial\,\mu_i}\left(\sum\limits_{i=1}^3\left(\frac{{
m}_{i}^2}{2\mu_i}+
\frac{\mu_i}{2}\right)\,+\,E_0(m_i,\mu_i)\right)\,=\,0,\end{equation}and
$C$ is the quark self-energy correction which is created by the
color magnetic moment of a quark propagating through the vacuum
background field \cite{qse}.
 The effect of the quark
self-energy is  to shift the mass spectrum by a global negative
amount~\footnote{Its negative sign is due to the paramagnetic
nature of the particular mechanism at work in this case.}:
\begin{equation} \label{self_energy}
C=-\frac{2\sigma}{\pi}\sum\limits_i\frac{\eta(t_i)}{\mu_i},
\,\,\,\,\,\,\,\, t_i\,=\,\frac{m_i}{T_g},\ee where $1/T_g$ is the
gluon correlation length. In this paper we use $T_g\,=$ 1~GeV.
 The function $\eta(t)$ is defined as
\be \eta(t)= t\int^\infty_0 z^2\, K_1(tz)\,
e^{-z}\,dz,\label{eq:eta} \ee where $K_1$ is the McDonald
function. The straightforward calculation yields \bea \eta(t)&
=&\,\,\,\,\,\,\, \frac{1+2t^2}{(1-t^2)^2}-
\frac{3t^2}{(1-t^2)^{5/2}}\, \ln\,
\frac{1+\sqrt{1-t^2}}{t},\,\,\,\,\,\,\,\,\,\, t<1, \nonumber\\
&&=\,\frac{1+2t^2}{(1-t^2)^2}-
\frac{3t^2}{(t^2-1)^{5/2}}\,\arctan\,
(\sqrt{t^2-1})\,\,\,\,\,\,\,\,\,\,t>1. \eea Note that
$\eta(0)\,=\,1$ and $\eta(t)\,\sim\,2/t^2$ as $t\,\to\,\infty$.
For the values of the bare strange quark mass $m_s\,=\,$ 100 MeV
and 175 MeV used in this paper $\eta_s\,=\,$0.9486 and 0.8882,
respectively.

\section{Hyperspherical formalism. Outline of the calculation.}
In this section, we briefly review the hyperspherical method,
which we use to calculate the masses of the ground and excited
hyperon states.

The baryon wave function depends on the three-body Jacobi
coordinates \begin{equation}\label{rho}
\bsl{\rho}_{ij}=\sqrt{\frac{\mu_{ij}}{\mu_0}}(\bsl{r}_i-\bsl{r}_j),
\end{equation}\begin{equation}\label{lambda}\bsl{\lambda}_{ij}=\sqrt{\frac{\mu_{ij,\,k}}{\mu_0}}
\left(\frac{m_i\bsl{r}_i+m_j\bsl{r}_j}{m_i+m_j}-\bsl{r}_k\right),\end{equation}
($i,j,k$ cyclic), where $\mu_{ij}$ and $\mu_{ij,k}$ are the
appropriate reduced masses
\begin{equation}\mu_{ij}=\frac{m_im_j}{m_i+m_j},~~
\mu_{ij,\,k}=\frac{(m_i+m_j)m_k}{m_i+m_j+m_k},\end{equation}
$\mu_0$ is an arbitrary parameter with the dimension of mass which
drops off in the final expressions. The coordinate
$\bsl{\rho}_{ij}$ is proportional to the separation of quarks $i$
and $j$ and coordinate $\bsl{\lambda}_{ij}$ is proportional to the
separation of quarks $i$ and $j$, and quark $k$. There are three
equivalent ways of introducing the Jacobi coordinates, which are
related to each other by linear transformations with the
coefficients depending on quark masses and Jacobian equal to
unity. In what follows we omit indices $i,j$.

In terms of the Jacobi coordinates the kinetic energy operator
$H_0$ is written as
\begin{equation} \label{H_0_jacobi} H_0= -\frac{1}{2\mu_0}
\left(\frac{\partial^2}{\partial\bsl{\rho}^2}
+\frac{\partial^2}{\partial\bsl{\lambda}^2}\right)
=-\frac{1}{2\mu}\left( \frac{\partial^2}{\partial
R^2}+\frac{5}{R}\frac{\partial}{\partial R}+
\frac{\bsl{L}^2(\Omega)}{R^2}\right), \end{equation} where $R$ is
the six-dimensional hyperradius,
\begin{equation} R^2=\bsl{\rho}^2+\bsl{\lambda}^2,\,\,\,\,\rho\,=\,R\,\sin\theta,\,\,\,\,\,
\lambda\,=\,R\,\cos\theta,\end{equation} $\Omega$ denotes five
residuary angular coordinates, and $\bsl{L}^2(\Omega)$ is an
angular operator
 \be{\bf
L}^2\,=\,\frac{\partial^2}{\partial
\theta^2}\,+\,4\cot\theta\,\frac{\partial}{\partial
\theta}-\frac{{\bf l}_{\rho}^2}{\sin^2\theta}\,-\,\frac{{\bf
l}_{\lambda}^2}{\cos^2\theta},\ee
whose eigenfunctions (the hyperspherical harmonics) satisfy
\begin{equation}
\label{eq: eigenfunctions} {\bf
L}^2(\Omega)\,Y_{[K]}(\Omega)_{\lambda})\,=\,-K(K+4)Y_{[K]}(\Omega),
\end{equation}
with $K$ being the grand orbital momentum.

In terms of $Y_{[K]}$ the wave function
$\psi(\bsl{\rho},\bsl{\lambda})$ can be written in a symbolical
shorthand as
\be\psi(\bsl{\rho},\bsl{\lambda})=\sum\limits_K\psi_K(R)Y_{[K]}(\Omega),\ee
where the set $[K]$ is defined by the the orbital momentum of the
state and the symmetry properties. We truncate this set using the
approximation $K\,=\,K_{\rm min}$ ($K_{\rm min}\,=\,0 $ for
$L\,=\,0$ and $K_{\rm min}\,=\,1 $ for $L\,=\,1$).
The corresponding hyperspherical harmonics are \be
Y_0(\Omega)\,=\,\sqrt{\frac{1}{\pi^3}},\,\,\,\,\,\,K\,=\,0,\ee
and\be\label{eq:harmonics}
\bsl{Y}_{\rho}(\Omega)\,=\,\sqrt{\frac{6}{\pi^3}}\,\frac{\bsl{\rho}}{R}\,,\,\,\,\,\,\,\,\,\,\,\,\,\,\,\,
\bsl{Y}_{\lambda}(\Omega)\,=\,\sqrt{\frac{6}{\pi^3}}\,\frac{\bsl{\lambda}}{R},\,\,\,\,\,\,K\,=\,1.\ee
The normalization coefficients in (\ref{eq:harmonics}) are easily
calculated using the relations \bea
&&\int\,\rho_i\rho_j\,f(\theta,\cos\chi)\,d\Omega\,=\,\frac{1}{3}\,\delta_{ij}\,
\int\,\rho^2\,f(\theta,\cos\chi)\,\,d\Omega,\,\nonumber\\
&&\int\,\lambda_i\lambda_j\,f(\theta,\cos\chi)\,d\Omega\,=\,\frac{1}{3}\,
\delta_{ij}\,\int\,\lambda^2\,f(\theta,\cos\chi)\,d\Omega,\eea
where $d\Omega\,=\,d{\bf n}_{\rho}\,d{\bf
n}_{\lambda}\sin^2\theta\cos^2\theta\,d\theta$.

For $\Lambda$ and $\Sigma$ we use the $uds$ basis in which the
strange quark is singled out as quark $3$ but in which the non
strange quarks are still antisymmetrized. In the same way, for the
$\Xi$ we use the $ssq$ basis with $q$ standing for $u$ or $d$
quarks, in which the non strange quark is singled out as quark
$3$.  The $uds$ basis states diagonalize the confinement problem
with eigenfunctions that correspond to separate excitations of the
non strange and strange quarks ($\bsl{\rho}$\,- and
$\bsl{\lambda}$\,- excitations, respectively). The nonsymmetrized
$uds$ or $ssq$ basis usually provides a much simplified picture of
the states. In particular, excitation of the $\bsl{\lambda}$
variable unlike excitation in $\bsl{\rho}$ involves the excitation
of the ``odd'' quark ($s$ for $\Lambda,\,\Sigma$ or $q$ for
$\Xi$).

We introduce the reduced function $u_{\gamma}(R)$ ($\gamma\,=\,0$
for $L\,=\,0$ and $\gamma\,=\,\rho$ or $\lambda$ for
$L\,=\,1$)~\footnote{In what follows we do not write explicitly
the magnetic quantum numbers of the vector hyperspherical
harmonics.}
\be\Psi_{\gamma}(R,\Omega)\,=\,\frac{u_{\gamma}(R)}{R^{5/2}}\cdot{
Y}_{\gamma}(\Omega), \ee and average the interaction $U=V_C+
V_{\text{string}}$ over the six-dimensional sphere $\Omega$ with
the weight $|Y_{\gamma}(\Omega)|^2$. Then one obtains the
Schr\"odinger equation for $u_{\gamma}(x)$
 \be\label{eq:se}\frac{d^2
u_{\gamma}}{dx^2}\,+\,2\left(E_0-\frac{(K+3/2)(K+5/2)}{2\,x^2}-V_{\gamma}(x)\right)u_{\gamma}(x)\,=\,0,\ee
where \be x\,=\,\sqrt{\,\mu_0}\,R\,.\ee In Eq. (\ref{eq:se}) \be
V_{\gamma}(x)\,=\, V_{\rm Y}^{\gamma}(x)\,+\,V_{\rm
Coulomb}^{\gamma}(x),\ee where\be\label{string} V_{\rm
Y}^{\gamma}(x)\,=\,\int \,|Y_{\gamma}(\Omega)|^2\,V_{\rm Y}({\bf
r}_1,\,{\bf r}_2,\,{\bf
r}_3)\,d\Omega\,=\,\sigma\,b_{\gamma}\,R\,=\,\sigma\, {\hat
b}_{\gamma}\,{x},\,\,\,\,\,\,\,\ee and \be\label{Coulomb} V_{\rm
Coulomb}^{\gamma}(x)\,=\,-\,\frac{2}{3}\,\alpha_s\,\int
\,|Y_{\gamma}(\Omega)|^2\,\sum\limits_{i<j}\,\frac{1}{r_{ij}}\,\,\,d\Omega\,=\,-\,\frac{2}{3}\,\alpha_s\,\frac{a_{\nu}}{R}
\,=\, -\,\frac{2}{3}\,\alpha_s\,\frac{{\hat
a}_{\gamma}}{x},\,\,\,\,\,\ee with \be{\hat
a}_{\gamma}\,=\,a\,\sqrt{\mu_0},\,\,\,\,\,\,\,\,\,\,{\hat
b}_{\gamma}\,=\,\frac{b_{\gamma}}{\sqrt{\mu_0}}\,.\ee In what
follows we denote
\be\label{eq:definition}\mu_1\,=\,\mu_2\,=\mu,\,\,\,\,\, {\rm
and}\,\,\,\,\, \mu_3\,=\,\kappa\,\mu\,.\ee The straightforward
analytical calculation of the integrals in Eq. (\ref{Coulomb})
yields \be {\hat
a}_{0}\,=\,\frac{32\,\sqrt{\mu}}{9\pi}\,\left(\frac{1}{\sqrt{2}}\,+\,2\,\sqrt{\frac{\kappa}{1+\kappa}}\right),\ee\be
{\hat
a}_{\rho}\,=\,\frac{32\,\sqrt{\mu}}{15\,\pi}\left(\,\sqrt{2}\,+\,\sqrt{\frac{\kappa}{\kappa+1}}\,\frac{5\kappa+6}{\kappa\,+\,1}
\right),\ee \be {\hat
a}_{\lambda}\,=\,\frac{32\,\sqrt{\mu}}{5\,\pi}\left(\frac{1}{\sqrt{2}}\,+\,\frac{1}{3}\,\sqrt{\frac{\kappa}{1\,+\,\kappa}}\,
\frac{4\,+\,5\kappa}{1\,+\,\kappa}\right),\ee while the
corresponding expressions for the string potential are more
complicated. We relegate the details of the numerical procedure
used to calculate the ${\hat b}_{\gamma}$ in Eq. (\ref{string}) to
Appendix.

For the doublet spin states one can introduce another possible
basis  corresponding to the fact that the $P$-wave hyperons
contain both positive and negative parity two-quark subsystems.
New basis states are given by the linear combinations
\be\label{eq: symmetrical basis}
\xi_s\,=\,\frac{\chi_a\,(12,3)\,{\bf
Y}_{\rho}\,+\,\chi_s\,(12,3)\,{\bf
Y}_{\lambda}}{\sqrt{2}}\,,\,\,\,\,\,\,\,\,\xi_a\,=\,\frac{\chi_a\,(12,3)\,{\bf
Y}_{\lambda}\,-\,\chi_s\,(12,3)\,{\bf Y}_{\rho}}{\sqrt{2}}\,,\ee
where $\chi_a$ and $\chi_s$ are the doublet spin functions which
are even and odd under the permutation of quarks 1 and 2,
respectively. In the $SU(3)$ limit $\mu_1\,=\,\mu_2\,=\,\mu_3$ the
spin-angular functions $\xi_s$ ($\xi_a$) are totally symmetric
(antisymmetric). Using basis states (\ref{eq: symmetrical basis})
one has for the P-wave matrix elements \be\label{Coulomb_1} V_{\rm
Coulomb}(x)\,=\,
-\,\frac{2}{3}\,\alpha_s\,\frac{\hat a}{x}\,,\,\,\,\,\,\,\, V_{\rm
Y}(x)\,=\,
\sigma\,{\hat b}\,{x}\,, \ee where \be\label{eq: new a and b}
{\tilde a}\,=\,\frac{{\tilde a}_{\rho}\,+\,{\tilde
a}_{\lambda}}{2}\,,\,\,\,\,\,{\tilde b}\,=\,\frac{{\tilde
b}_{\rho}\,+\,{\tilde b}_{\lambda}}{2}\,.\ee

\section{RESULTS}
In this Section we present the results obtained for $qqq$, $qqs$
and $ssq$ baryons. As already stated, we disregard the hyperfine
interactions which give spin-doublet~-~spin-quartet splittings and
the spin-orbit interactions, which describes the fine structure of
states.
Note that with its attractive $\delta$-function, the hyperfine
interaction produces effects which are must stronger than those
one would obtain from the lowest order perturbation theory, in
which the $\Lambda$ and $\Sigma$ hyperons are almost degenerate.
The large hyperfine effects in $\Delta\,-\, N$ or
$\Sigma\,-\,\Lambda$ splittings are usually described by the
smeared $\delta$-function \cite{CI86} and/or meson exchanges
between quarks (see {\it e.g.} \cite{VGV05}), and generally
require additional model-dependent assumptions about the structure
of interquark forces.

We do not perform a systematic study in order to determine the
best set of parameters $\sigma$, $\alpha_s$ and $m_s$ to fit the
hyperon spectra. Instead, we employ some typical values of
$\sigma$ and $\alpha_s$ that have been used for the description of
the ground state baryons: the string tension $\sigma$ is taken to
be commonly used value of 0.15 GeV$^2$ and the strong coupling
constant $\alpha_s\,=\,$ 0.39. For the nucleon we slightly varied
$\alpha_s$ to illustrate the sensitivity of the results to the
chosen input. In our calculations we use the values of the current
light quark masses ${m}_u\,=\,{ m}_d\,=\,7\,$ MeV, ${m}_s\,=\,$
100 and 175 MeV.

We begin with the discussion of the $qqq$ states with
$L^P\,=\,0^+$ and $1^-$. For $L\,=\,1$ we use the spin-angular
functions (\ref{eq: symmetrical basis}). In Table
\ref{tab:nucleon} we display the nucleon masses for the three
choices of $\alpha_s$: 0.39, 0.5 and 0.6~ \footnote{The results
for $L^P\,=\,0^+$ state and $\alpha_s\,=\,$0.39, as those for the
$qqs$ and $sss$ states with $m_s\,=\,175$ MeV (shown in Table
\ref{tab:different m_s}) have been previously obtained in
\cite{NT}, and we quote the (slightly updated) results here for
comparison.}. The last value have been used in the Capstik-Isgur
model \cite{CI86}. Increasing $\alpha_s$ by $\sim\,0.1$ decreases
the nucleon mass by $\sim\,50$ MeV. We get
$\frac{1}{2}\,(N\,+\,\Delta)_{\rm theory}\,=\,$1228, 1181 and 1131
MeV for $L\,=\,0$ and $\alpha_s\,=\,$0.39, 0.5 and 0.6,
respectively, {\it vs}  $\frac{1}{2}\,(N\,+\,\Delta)_{\rm
exp}\,=\,$ 1085 MeV.
For $L\,=\,1$ we obtain 1770 MeV, 1653 MeV and 1613 MeV,
respectively. The excitation energies only weekly depend on
$\alpha_s$: cost (in $\Delta L$) is 469 MeV for
$\alpha_s\,=\,0.39$, 472 MeV for $\alpha_s\,=\,0.5$, and 482 MeV
for $\alpha_s\,=\,0.6$.

The $qqs$ and $ssq$ states which belong to the octet are
structurally identical to the nucleon: only one and two,
respectively, of the light quarks are replaced by a strange quark.
Consequently, the analysis of these states and the results are
only a variation of what has been found for the nucleon system.
Table \ref{tab:different m_s} displays the sensitivity of hyperon
masses to the chosen value of ${m}_s$.  Increasing $m_s$ by 75 MeV
increases $\mu_s$ by 30 MeV both for $qqs$ and  $ssq$, but
practically does not affect $\mu_q$. As the result the masses of
the $qqs$ and $ssq$ states are increased by 40 MeV and 75 MeV,
respectively. We get
$\frac{1}{4}(\Lambda\,+\,\Sigma\,+\,2\,\Sigma^*)_{\rm
theory}\,=\,$ 1278 and 1317 MeV for $m_s\,=\,$ 100 and 175 MeV,
respectively, {\it vs}
$\frac{1}{4}(\Lambda\,+\,\Sigma\,+\,2\,\Sigma^*)_{\rm exp}\,=\,$
1267 MeV.

Table \ref{tab:rho and lambda} displays our main results for  the
$\rho$ and $\lambda$ excitations of the $P$ wave $qqs$ and $ssq$
hyperons. As in Table  \ref{tab:different m_s} we present here the
dynamical quark masses, zero-order eigenenergies and the hyperon
masses for $m_s\,=\,$ 100 and 175 MeV. The dynamical quark masses
$\mu_i$ corresponding to the excitations $\rho$ and $\lambda$ are
somewhat different. This is not surprising because these
quantities can be considered as the average kinetic energies of
the current quarks, which are larger for the quarks in the P-wave
and smaller for the quarks in the S-wave. The eigenenergies $E_0$
of the $\rho$ and $\lambda$ excitations are degenerate, the masses
of the corresponding states
 are nearly degenerate: they differ no more than 10 MeV~\footnote{This degeneracy
 disappear for the heavy baryons, in which case the masses of the $\rho$ and
$\lambda$ excitations differ by $\sim$ 100 MeV \cite{NV}.}. This
difference is due to the difference of the dynamical masses
$\mu_i$ which enter the mass formula (\ref{M_B}) and  the self
energy contribution (\ref{self_energy}).

In Table \ref{tab:hyperons} we present the zero-order $qqs$ and
$ssq$ masses calculated for the mixture of the $\rho$ and
$\lambda$ excitations, Eq. (\ref{eq: symmetrical basis}), for
$m_s\,=\,$ 100 MeV. Comparing the results with those of Table
\ref{tab:rho and lambda} we observe again the same eigenenergies
$E_0$ and slightly different hyperon masses $M$. The excitation
energies
 $\Delta\,=\,M(L=1)\,-\,M(L=0)$ are of the order of 460 MeV both for $\Sigma$
 and $\Xi$ and also coincide with the excitation energies for the
 nucleon.

The discrepancy of the results shown in Tables
\ref{tab:nucleon}-\ref{tab:hyperons} with the experimental ones
hints that it is the chiral physics, missing our approach, that
could shift nucleon and $\Lambda$ states down \cite{GR96b}.
However, the chiral effects are less important for the $\Sigma$
states. Therefore we identify two approximately degenerate $qqs$
excited states of negative parity with the P-wave $\Sigma$
resonances. The PDG \cite{PDG06} lists $D_{13}$, $S_{11}$ and
$D_{15}$ resonances with $I\,=\,1$, $J^P\,=\,(3/2)^-$,~
$\,(1/2)^-$, $\,(5/2)^-$ and masses 1670, 1750 and 1775 MeV,
respectively. The latter state corresponds to the $\lambda$ -
excitation with $S=3/2$. The results are compatible with the known
states, showing discrepancies with the experimental data of order
5\% or less.

We finally remark that our result for the negative parity ground
state in the $\Xi$ channel $M(L\,=\,1)\,=\,$ 1781 MeV exactly
agrees with the recent finding from the lattice quenched
calculation \cite{B06}. The other theoretical predictions for this
state are listed in Table \ref{tab:xi theory}.

\setcounter{equation}{0}
\def\theequation{A.\arabic{equation}}

\section{Conclusions}
In this paper we have extended  our previous study of the ground
states of the $qqq$, $qqs$ and $ssq$ baryons to the description of
their first angular excitations. We use the EH method.  The
three-quark problem has been solved using the hyperspherical
approach. For each baryon we have calculated the dynamical quark
masses $\mu_i$ from Eq. (\ref{partial}), energy eigenvalues $E_0$
from Eq. (\ref{eq:se}), and the baryon masses (\ref{M_B}) with the
self-energy corrections (\ref{self_energy}). The main results are
given in Tables \ref{tab:rho and lambda} and \ref{tab:hyperons}.
 Our study suggests that a good description of
the P-wave baryons can be obtained with a spin independent energy
eigenvalues corresponding to the confinement plus Coulomb
potentials.
Moreover
this comparative study gives a better insight into the quark model
results where the constituent masses encode the QCD dynamics.

The authors are grateful to M.A.Trusov for useful remarks. This
work was supported by RFBR grants 05-02-17869 and 06-02-17120.

\section*{Appendix. The string junction potential in the hyperspherical formalism}

Recall the definition of the minimal length string Y--shaped
configuration \cite{CI86}. Let $\varphi_{ijk}$ be the inner angle
between the line from quark $i$ to quark $j$ and that from quark
$j$ to quark $k$. One should distinguish two cases. If all the
inner angles of the triangle formed by three constituent quarks
sitting at the apexes are smaller than $120^\circ$, the junction
point coincides with the so-called Torrichelli point of the
triangle and
\bea &&r_{\rm
min}=\\\label{eq:r
min_a}&&\left(\frac{1}{2}\sum\limits_{i<j}r_{ij}^2
 +\frac{\sqrt{3}}
{2}\sqrt{(r_{12}+r_{31}+r_{23})(-r_{12}+r_{31}+r_{23})(r_{12}-r_{31}+r_{23})(r_{12}+r_{31}-r_{23})}
\right)^{1/2}\nonumber \eea The relative distances $r_{ij}$ in
(A.1) are expressed in terms of the Jacoby coordinates:

\begin{equation}
\begin{aligned}
r_{12}\,=\,& \sqrt{\frac{2\mu_0}{\mu}}\,\rho\,,
\\
r_{31}\,=\,&
\sqrt{\frac{\mu_0}{2\mu}}\left(\rho^2\,+\,\frac{\kappa\,+\,2}{\kappa}\,\lambda^2\,+\,
2\,\sqrt{\frac{\kappa+2}{\kappa}}\,\rho\,\lambda\,\cos\chi\right)^{1/2},\\
r_{23}\,=\,&
\sqrt{\frac{\mu_0}{2\mu}}\left(\rho^2\,+\,\frac{\kappa\,+\,2}{\kappa}\,\lambda^2\,-\,
2\,\sqrt{\frac{\kappa+2}{\kappa}}\,\rho\lambda\,\cos\chi\right)^{1/2},
\end{aligned}
\label{r_123}
\end{equation}
where $\cos\chi\,=\,{\bf n}_{\rho}{\bf n}_{\lambda}$. Substituting
these expressions into (A.1), one obtains

\be \label{l_0} r_{\rm
min}\,=\,x\,l_0(\theta,\chi)\,=\,\frac{x}{\sqrt{\mu}}\,
\left(\frac{3}{ 2}\,\sin^2\theta\,+\,
\sqrt3\,\,\sqrt{\frac{\kappa\,+\,2}{\kappa}}\sin\theta\cos\theta\sin\chi
\,+\,\frac{\kappa\,+\,2}{2\,\kappa}\cos^2\theta\right)^{\frac{1}{2}}\,.
\ee If $\varphi_{ijk}$ is equal to or greater than $2\pi/3$, the
lowest energy configuration has the junction at the apex connected
with quark $j$: \be r_{\rm min}\,=\,r_{ij}\,+\,r_{jk},\label{eq:r
min b}\ee where \be r_{ij}\,=\,x\,l_{ij}(\theta,\chi).\ee
Accordingly, in the $\theta-\chi$ plane one should distinguish the
four regions:
\begin{enumerate}
\item[(i)]region I: $\cos\varphi_{ijk}\geq -\,1/2$,
$r_{\rm min}\,=\,x\,l_0(\theta,\chi)$, where $l_0(\theta,\chi)$ is
defined in (\ref{l_0}), \item[(ii)] region II:
$\cos\varphi_{312}\leq -\,1/2$,\, $\chi\,\geq\,\,\frac{2\pi}{3}$,
$r_{\rm
min}\,=\,r_{12}\,+\,r_{31}\,=\,x(l_{12}(\theta)\,+\,l_{31}(\theta,\chi))$,
\item[(iii)]region III: $\cos\varphi_{123}\leq -\,1/2$,\,$\chi\leq
\,\,\frac{\pi}{3}$\,
$r_{\rm
min}\,=\,r_{12}\,+\,r_{23}\,=\,x(l_{12}(\theta)\,+\,l_{23}(\theta,\chi))$,
and \item[(iv)]region IV: $\cos\varphi_{231}\leq -\,1/2$,
$r_{\rm
min}\,=\,r_{31}\,+\,r_{23}\,=\,x(l_{31}(\theta,\chi)\,+\,l_{23}(\theta,\chi))$.
\end{enumerate}
The expressions for $l_{12}(\theta)$, $l_{31}(\theta,\chi)$ and
$l_{23}(\theta,\chi)$ follow from Eqs. (\ref{r_123}). The
boundaries between region I and regions II, III are given by
\be\theta_{1,2}(\chi)\,=\,
\arctan\left(\sqrt{\frac{\kappa\,+\,2}{\kappa}}(\,\mp\,\cos\chi\,-\,\frac{1}{\sqrt{3}}
\sin\chi)\right),\ee while the boundary between regions I and IV
is given by \be\theta_3(\chi)\,=\,
\arctan\left(\sqrt{\frac{1}{3}}\,\sqrt{\frac{\kappa\,+\,2}{\kappa}}\,(\,\sin\chi\,
+\,\sqrt{\sin^2(\chi)\,+\,3}\,\,)\,\right).\ee The constants
${\hat b}_{\gamma}$ in Eq. (\ref{string}) are expressed in terms
of the four integrals:

\be {\hat
b}_{\gamma}\,=\,\frac{1}{x}\,\left(\int\limits_I\,+\,\int\limits_{II}\,+\int\limits_{III}\,+\int\limits_{IV}
\right)\,r_{\rm min}\,d\Omega_{\gamma},\ee or, in the explicit
form,
\begin{eqnarray}
&&{\hat b}_{\gamma}\,=
\,\int\limits_0^{\frac{\pi}{3}}\sin\chi\,d\chi\,\times
\left(\int\limits_0^{\theta_2(\chi)}(l_{12}(\theta,\chi)\,+
\,l_{23}(\theta,\chi))\,d\,\Omega_{\gamma}\,+\,\right.\\
&&\left.
+\,\int\limits_{\theta_2(\chi)}^{\theta_3(\chi)}l_0(\theta,\,\chi)
\,d\,\Omega_{\gamma}(\theta)\,+\,\int\limits_{\theta_3(\chi)}^{\frac{\pi}{2}}
(l_{23}(\theta,\chi)\,+l_{31}(\theta,\chi)\,)\,d\,\Omega_{\gamma}\,\right)\nonumber\\
&&+\,
\int\limits_{\frac{\pi}{3}}^{\frac{2\pi}{3}}\sin\chi\,d\chi\,
\,\left(\int\limits_0^{\theta_3(\chi)}\,l_0(\theta,\chi)\,d\,\Omega_{\gamma}\,
+\,\int\limits_{\theta_3(\chi)}^{\frac{\pi}{2}}\,(l_{23}(\theta,\chi)\,+\,l_{31}(\theta,\chi))\,
d\,\Omega_{\gamma}\,\right)\,+\,
\nonumber\\
&&+\,\int\limits_{\frac{2\pi}{3}}^{\pi}\sin\chi\,d\chi\,\times\nonumber
\left(\int\limits_0^{\theta_1(\chi)}(l_{12}(\theta,\chi)\,+
\,l_{31}(\theta,\chi))\,d\,\Omega_{\gamma}(\theta)\,+\,\right.\\
&&\left.
+\,\int\limits_{\theta_1(\chi)}^{\theta_3(\chi)}l_0(\theta,\,\chi)
\,d\,\Omega_{\gamma}\,+\,\int\limits_{\theta_3(\chi)}^{\frac{\pi}{2}}
(l_{23}(\theta,\chi)\,+l_{31}(\theta,\chi)\,)\,d\,\Omega_{\gamma}\,\right),\nonumber\label{eq:
b for strings}
\end{eqnarray}
where
\be
d\Omega_0\,=\,\frac{8}{\pi}\sin^2\theta\cos^2\theta\,\sin\chi\,d\chi
d\theta\ee for $K\,=\,0,$  \be d\Omega_{\rho}\,=\,\frac{1}{3}\cdot
\frac{48}{\pi}\,\sin^4\theta\cos^2\theta\sin\chi\,d\chi
d\theta,\,\,\,\,\,\,\,\,\,\,\,\,
d\Omega_{\lambda}\,=\,\frac{1}{3}\cdot
\frac{48}{\pi}\,\sin^2\theta\cos^4\theta\sin\chi\,d\chi d\theta\ee
for $K\,=\,1$.

\newpage

\begin{table}[t] \centering
\begin{tabular}{cccccccccc|} \hline\hline\\
$\alpha_s$& $L$ &$\mu$&$E_0$&$M_N$&$\Delta$ \\
\\
\hline\hline
\\
0.39&0&408&1318&1228&
\\
0.5&0&425&1217&1181&
\\
0.6&0&442&1121&1131&
\\ \\
\hline \\  0.39& 1& 457&1638&1697&469
\\
 0.5& 1&469&1560&1653&472
 \\
 0.6 &1&481&1487&1613&482
  \\ \\
\hline\hline
\end{tabular}
\caption{Ground and excited state nucleon masses for
$\alpha_s\,=\,$ 0.39, 0.5 and 0.6. For each case shown are the
dynamical quark masses $\mu$, defined by Eq. (\ref{partial}),
eigenenergies $E_0$ of Eq. (\ref{eq:se}), the nucleon masses given
by Eq. (\ref{M_B}), and $\Delta\,=\,M(L=1)\,-\,M(L=0)$ (all in
units of MeV). P-wave eigenenergies and nucleon masses correspond
to the spin angular functions (\ref{eq: symmetrical basis}).
} \label{tab:nucleon}
\end{table}
\begin{table}[t] \centering
\begin{tabular}{cccccc} \hline\hline\\
 & ${m}_s$&$\mu_1\,=\,\mu_2$&$\mu_3$&$E_0$&$M$ \\ \\
\hline \\
$qqs$ &100&410&424&1308&1278\\
&175&414&453&1291&1317\\ \\
\hline\\
$ssq$ &100&426&412&1298&1327\\
&175&458&419&1266&1402\\ \\
 \hline\hline
\end{tabular}
\caption{Ground state hyperons for ${m}_s\,=\,$ 175 and 100 MeV.
The notations are the same as in Table \ref{tab:nucleon}.}
\label{tab:different m_s}
\end{table}
\begin{table}[t] \centering
\begin{tabular}{ccccccc} \hline\hline\\
Hyperon&$m_s$ &Excitation& $\mu_1\,=\,\mu_2$ & $\mu_3$& $E_0$& $M$\\
\\ \hline\\
$qqs$&100 &$\rho$&479&431&1627&1724\\
&100&$\lambda$&438&509&1629&1717\\
& 175&$\rho$  & 482 & 457 & 1612 & 1774\\
&175& $\lambda$&440 &532 & 1616& 1758\\
\\ \hline \\
$ssq$&100&$\rho$&491&419&1621&1745\\
&100&$\lambda$&452&500&1620&1752\\
&175&  $\rho$  & 518 & 423 & 1594& 1829\\
&175& $\lambda$ &480 &506&1591&1845\\
\\
\hline\hline
\end{tabular}
\\
\caption{Masses of the $\rho$ and $\lambda$  hyperon excitations.
Shown are the dynamical quark masses $\mu_i$, the confinement
energies $E_0$ and the hyperon masses $M$ (all in units of MeV).
} \label{tab:rho and lambda}
\end{table}
\begin{table}[t] \centering
\begin{tabular}{cccccccccc|} \hline\hline\\
Hyperon & $L^P$ &$\mu_1\,=\,\mu_2$&$\mu_3$&$E_0$&$M$&$\Delta$ \\
\\
\hline\hline \\
qqs &$0^+$&410&424&1308&1278&\\
&$1^-$& 458&471&1630&1739&461 \\ \\
\hline\\
ssq&$0^+$&426&412&1298&1327&&\\
&$1^-$&472&460&1622&1781&454\\ \\
\hline\hline
\end{tabular}
\caption{Solutions of Eqs. (\ref{M_B}), (\ref{eq:se}) for the
hyperon states with $L\,=\,0,1$. $\alpha_s\,=\,$ 0.39, ${
m}_s\,=\,$ 100 MeV.  P-wave eigenenergies and the hyperon  masses
correspond to the spin-angular functions (\ref{eq: symmetrical
basis}).} \label{tab:hyperons}
\end{table}
\begin{table*}[t]
\centering
\begin{tabular}{c|ccccccccc} \hline\hline\\
State & \cite{CIK81} & \cite{CI86} & \cite{GR96b} &  \cite{CC00} &
\cite{BIL00} &\cite{LL02} (\cite{JO96})&\cite{Oh07}&This work&PDG
\cite{PDG06}
\\ \hline \\
$\Xi(\frac12^-)$ & $1785$ & $1755$ & $1758$ & $1780$ & $1869$ &
$1550$ ($1630)$&1660&1781& $\Xi(1690)\,?$\\  $\Xi(\frac32^-)$ &
$1800$ & $1785$ & $1758$ & $1815$ & $1828$ & $1840$
&1820&$1781$&$\Xi(1820)\,$\\ \\ \hline
\end{tabular}
\caption{Low-lying $\Xi$  spectrum of spin $L\,=\,1$ predicted by
the non-relativistic quark model of Chao, Isgur and Karl
\cite{CIK81}, relativized quark model of Capstick and Isgur
\cite{CI86}, Glozman-Riska model \cite{GR96b}, large $N_c$
analysis \cite{CC00}, algebraic model \cite{BIL00}, QCD sum rules
\cite{LL02}, and the Skirm model \cite{Oh07}. The question mark in
the last column means that the $J^P$ quantum numbers are not
identified by PDG. The mass is given in the unit of MeV.}
\label{tab:xi theory}
\end{table*}
\end{document}